\documentclass{article}
\usepackage{ismir,amsmath,cite,url}
\usepackage{graphicx}
\usepackage{color}
\usepackage{bm}

\usepackage{makecell}

\usepackage{csvsimple}

\title{A database linking piano and orchestral \textit{MIDI} scores with application to automatic projective orchestration}

\multauthor
{L\'eopold Crestel$^1$ \hspace{1cm} Philippe Esling$^1$} { \bfseries{Lena Heng$^2$ \hspace{1cm} Stephen McAdams$^2$ \hspace{1cm}}\\
$^1$ Music Representations, IRCAM, Paris, France\\
$^2$ Schulich School of Music, McGill University, Montr\'eal, Canada\\
{\tt\small leopold.crestel@ircam.fr}
}
\def\authorname{L\'eopold Crestel, Philippe Esling, Lena Heng, Stephen McAdams}

\begin{document}

\maketitle
\begin{abstract}
This article introduces the Projective Orchestral Database (\textit{POD}), a collection of \textit{MIDI} scores composed of pairs linking piano scores to their corresponding orchestrations. To the best of our knowledge, this is the first database of its kind, which performs piano or orchestral prediction, but more importantly which tries to learn the correlations between piano and orchestral scores.
Hence, we also introduce the projective orchestration task, which consists in learning how to perform the automatic orchestration of a piano score. 
We show how this task can be addressed using learning methods and also provide methodological guidelines in order to properly use this database.
\end{abstract}

\section{Introduction}
\begin{figure}
\centering
\includegraphics[scale=0.14]{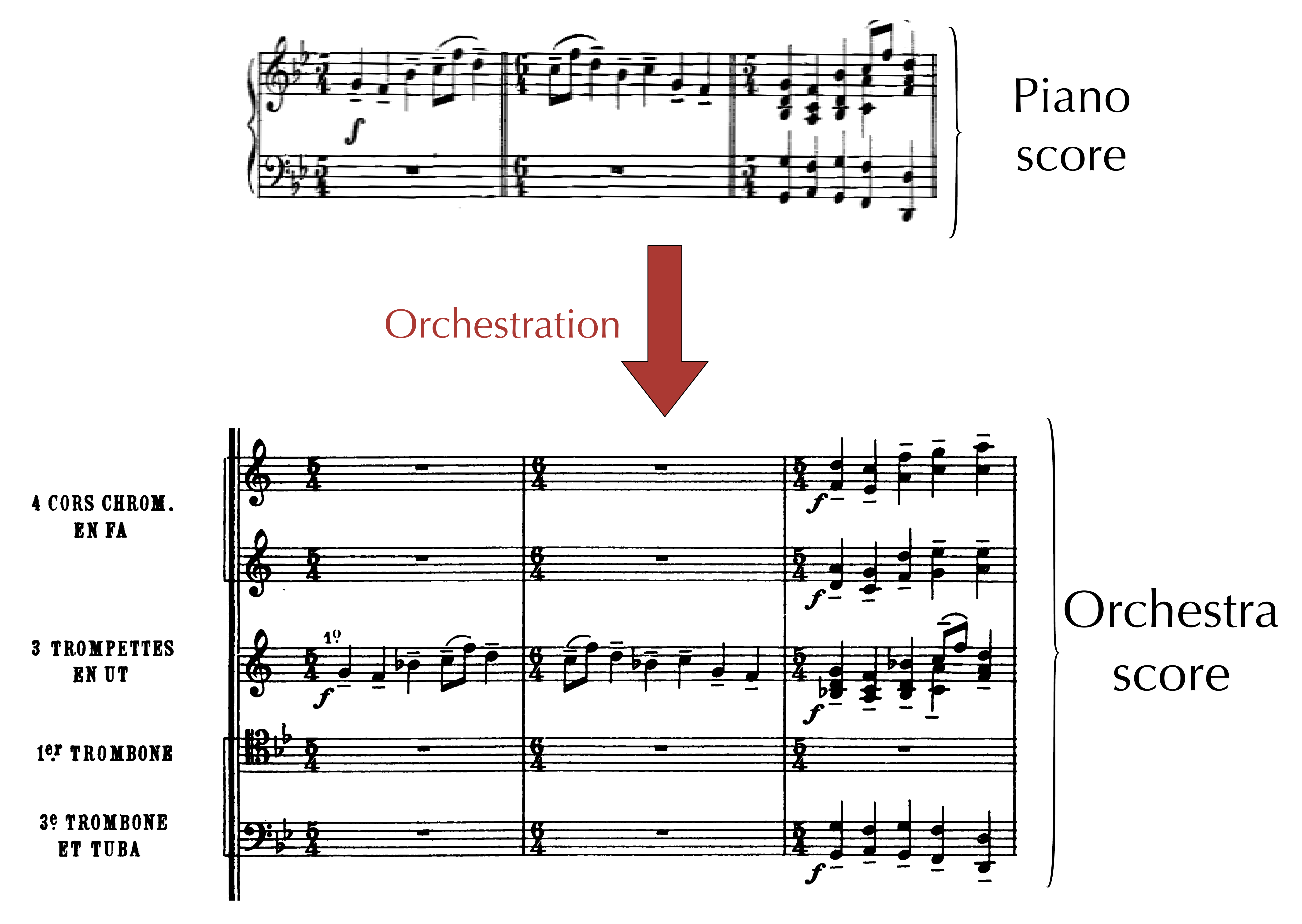}
\caption{\textit{Projective orchestration} of the first three bars of Modest Mussorgsky's piano piece \textit{Pictures at an Exhibition} by Maurice Ravel. Piano notes are assigned to one or several instruments, possibly with doubling or harmonic enhancement.}
\label{fig:orch}
\end{figure}

Orchestration is the subtle art of writing musical pieces for the orchestra by combining the properties of various instruments in order to achieve a particular musical idea \cite{koechli_orch,Rimsky-Korsakov:1873aa}. 
Among the variety of writing techniques for orchestra, we define as \textit{projective orchestration} \cite{esling2010dynamic} the technique which consists in first writing a piano score and then orchestrating it (akin to a projection operation, as depicted in \figref{fig:orch}).
This technique has been used by classic composers for centuries. One such example is the orchestration by Maurice Ravel of \textit{Pictures at an Exhibition}, a piano work written by Modest Mussorgsky.
This paper introduces the first dataset of musical scores dedicated to projective orchestrations.
It contains pairs of piano pieces associated with their orchestration written by famous composers. 
Hence, the purpose of this database is to offer a solid knowledge for studying the correlations involved in the transformation from a piano to an orchestral score.

  The remainder of this paper is organized as follows. First, the motivations for a scientific investigation of orchestration are exposed (section 2). By reviewing the previous attempts, we highlight the specific need for a symbolic database of piano and corresponding orchestral scores. 
In an attempt to fill this gap, we built the \textit{Projective Orchestral Database} (\textit{POD}) and detail its structure in section 3. In section 4, the automatic projective orchestration task is proposed as an evaluation framework for automatic orchestration systems. 
We report our experiment with a set of learning-based models derived from the Restricted Boltzmann Machine \cite{taylor2009factored} and introduce their performance in the previously defined evaluation framework. Finally, in section 5 we provide methodological guidelines and conclusions.

\section{A scientific investigation of orchestration}
Over the past centuries, several treatises have been written by renowned composers in an attempt to decipher some guiding rules in orchestration \cite{koechli_orch,piston-orch,Rimsky-Korsakov:1873aa}.
Even though they present a remarkable set of examples, none of them builds a systemic set of rules towards a comprehensive theory of orchestration.
The reason behind this lack lies in the tremendous complexity that emerges from orchestral works. 
A large number of possible sounds can be created by combining the pitch and intensity ranges of each instruments in a symphonic orchestra. 
Furthermore, during a performance, the sound produced by a mixture of instruments is also the result of highly non-linear acoustic effects. 
Finally, the way we perceive those sounds involves complex psychoacoustic phenomena \cite{lembke2012timbre,tardieu2012perception,mcadams2009perception}.
It seems almost impossible for a human mind to grasp in its entirety the intertwined mechanisms of an orchestral rendering.

Hence, we believe that a thorough scientific investigation could help disentangle the multiple factors involved in orchestral works. 
This could provide a first step towards a greater understanding of this complex and widely uncharted discipline.
Recently, major works have refined our understanding of the perceptual and cognitive mechanisms specifically involved when listening to instrumental mixtures \cite{pressnitzer2000perception,tardieu2012perception,mcadams2013timbre}. Orchids, an advanced tool for assisting composers in the search of a particular sonic goal has been developed \cite{esling2010dynamic}. It relies on the multi-objective optimization of several spectro-temporal features such as those described in \cite{peeters2011timbre}.

However, few attempts have been made to tackle a scientific exploration of orchestration based on the study of musical scores.
Yet, symbolic representations implicitly convey high-level information about the spectral knowledge composers have exploited for timbre manipulations.
In \cite{cookerly2010complete} a generative system for orchestral music is introduced. Given a certain style, the system is able to generate a melodic line and its accompaniment by a full symphonic orchestra. Their approach relies on a set of templates and hand-designed rules characteristic of different styles.
\cite{Pachet:2016:JOA:3004291.2897738} is a case study of how to automatically transfer the \textit{Ode to joy} to different styles. Unfortunately, very few details are provided about the models used, but it is interesting to observe that different models are used for different styles.
Automatic arrangement, which consists in reducing an orchestral score to a piano version that is can be played by a two-hand pianist, has been tackled in \cite{huang2012towards} and \cite{automatic_arranging_smc}.
The proposed systems rely on an automatic analysis of the orchestral score in order to split it into structuring elements. Then, each element is assigned a role which determines whether it is played or discarded in the reduction.
To the best of our knowledge, the inverse problem of automatically orchestrating a piano score has never been tackled.
However, we believe that unknown mechanisms of orchestration could be revealed by observing how composers perform projective orchestration, which essentially consists in highlighting an existing harmonic, rhythmic and melodic structure of a piano piece through a timbral structure.

Even though symbolic data are generally regarded as a more compact representation than a raw signal in the computer music field, 
the number of pitch combinations that a symphonic orchestra can produce is extremely large. 
Hence, the manipulation of symbolic data still remains costly from a computational point of view.
Even through computer analysis, an exhaustive investigation of all the possible combinations is not feasible.
For that reason, the approaches found in the literature rely heavily on heuristics and hand-designed rules to limit the number of possible solutions and decrease the complexity.
However, the recent advents in machine learning have brought techniques that can cope with the dimensionality involved with symbolic orchestral data.
Besides, even if a wide range of orchestrations exist for a given piano score, all of them will share strong relations with the original piano score.
Therefore, we make the assumption that projective orchestration might be a relatively simple and well-structured transformation lying in a complex high-dimensional space.
Neural networks have precisely demonstrated a spectacular ability for extracting a structured lower-dimensional manifold from a high-dimensional entangled representation \cite{LeCun:2015aa}.
Hence, we believe that statistical tools are now powerful enough to lead a scientific investigation of projective orchestration based on symbolic data.

These statistical methods require an extensive amount of data, but there is no symbolic database dedicated to orchestration.
This dataset is a first attempt to fill this gap by building a freely accessible symbolic database of piano scores and corresponding orchestrations.

\section{Dataset}
\subsection{Structure of the Database}
  The database can be found on the companion website
  \footnote{\url{https://qsdfo.github.io/LOP/database}}
  of this article, along with statistics and Python code for reproducibility.

\subsubsection{Organization}
The Projective Orchestral Database (\textit{POD}) contains 392 \textit{MIDI} files.
Those files are grouped in pairs containing a piano score and its orchestral version.
Each pair is stored in a folder indexed by a number.
The files have been collected from several free-access databases \cite{imslp} or created by professional orchestration teachers.
	
\subsubsection{Instrumentation}
As the files gathered in the database have various origins, different instrument names were found under a variety of aliases and abbreviations.
Hence, we provide a comma-separated value (\textit{CSV}) file associated with each \textit{MIDI} file in order to normalize the corresponding instrumentations. In these files, the track names of the \textit{MIDI} files are linked to a normalized instrument name.

\subsubsection{Metadata}
For each folder, a CSV file with the name of the folder contains the relative path from the database root directory, the composer name and the piece name for the orchestral and piano works.
A list of the composers present in the database can be found in table \ref{tab:composers}.
It is important to note the imbalanced representativeness of composers in the database.
It can be problematic in the learning context we investigate, because a kind of stylistic consistency is \textit{a priori} necessary in order to extract a coherent set of rules.
Picking a subset of the database would be one solution, but another possibility would be to add to the database this stylistic information and use it in a learning system.

\begin{table}[ht!]
\centering
\scalebox{0.5}{
{\renewcommand\arraystretch{1.25}
	\begin{tabular}{|c|c|c|c|c|}
    	\hline
		\thead{Composer} & \thead{Number of\\piano files} & \thead{Percentage\\piano frames} & \thead{Number of\\orchestra files} & \thead{Percentage\\orchestra frames}
    	\begin{filecontents*}{stat_composer.csv}
composer,numpiano,percentagepiano,numorch,percentageorch
Arcadelt. Jacob,,,1,0.07
Arresti. Floriano,3,0.57,,
Bach. Anna Magdalena,3,0.43,,
Bach. Johann Sebastian,9,4.57,4,0.81
Banchieri. Adriano,,,1,0.32
Beethoven. Ludwig Van,1,0.60,38,42.28
Berlioz. Hector,,,1,0.14
Brahms. Johannes,,,3,0.28
Buxtehude. Dietrich,1,0.21,,
Byrd. William,1,0.13,,
Charpentier. Marc-Antoine,,,2,0.38
Chopin. Frederic,,,2,0.44
Clarke. Jeremiah,,,1,0.23
Debussy. Claude,1,0.59,6,0.90
Dvorak. Anton,,,6,2.42
Erlebach. Philipp Heinrich,,,1,0.10
Faure. Gabriel,,,1,0.60
Fischer. Johann Caspar Ferdinand,1,0.10,,
Gluck. Christoph Willibald,,,1,1.61
Grieg. Edvard,,,1,2.10
Guerrero. Francisco,1,0.12,,
Handel. George Frideric,4,1.00,1,0.75
Haydn. Joseph,,,6,1.01
Kempff. Wilhelm,1,1.58,,
Leontovych. Mykola,,,2,0.22
Liszt. Franz,34,39.98,,
Mahler. Gustav,,,1,0.85
Mendelssohn. Felix,,,2,1.41
Moussorgsky. Modest,,,1,0.04
Mozart. Wolfgang Amadeus,1,0.71,8,1.45
Okashiro. Chitose,3,1.09,,
Pachelbel. Johann,1,0.15,,
Praetorius. Michael,,,2,0.14
Purcell. Henry,,,1,0.08
Ravel. Maurice,6,6.49,8,6.69
Rondeau. Michel,2,0.25,1,0.14
Schonberg. Arnold,,,1,0.21
Schumann. Robert,,,1,0.05
Shorter. Steve,1,0.26,,
Smetana. Bedrich,,,1,0.61
Soler. Antonio,1,0.54,,
Strauss. Johann,,,1,0.04
Strauss. Richard,,,1,0.22
Stravinsky. Igor,,,4,0.94
Tchaikovsky. Piotr Ilyich,,,36,20.08
Telemann. Georg Philipp,,,2,1.04
Unknown. ,107,40.18,28,7.47
Vivaldi. Antonio,,,4,2.94
Walther. Johann Gottfried,1,0.14,,
Wiberg. Steve,,,1,0.75
Zachow. Friedrich Wilhelm,1,0.32,2,0.23
\end{filecontents*}
\csvreader[head to column names]{stat_composer.csv}{}
    	{\\ \hline \composer & \numpiano & \percentagepiano & \numorch & \percentageorch }
        \\\hline
	\end{tabular}}
    }
\caption{This table describes the relative importance of the different composers present in the database. For each composer, the number of piano (respectively orchestral) scores in the database are indicated in the second (respectively fourth) column. The total number of files is 184 x 2 = 392.
As the length of the files can vary significantly, a more significant indicator of a composer's representativeness in the database is the ratio of the number of frames from its scores over the total number of frames in the database.}
\label{tab:composers}
\end{table}

\begin{figure}[t!]
\centering
\includegraphics[scale=0.45]{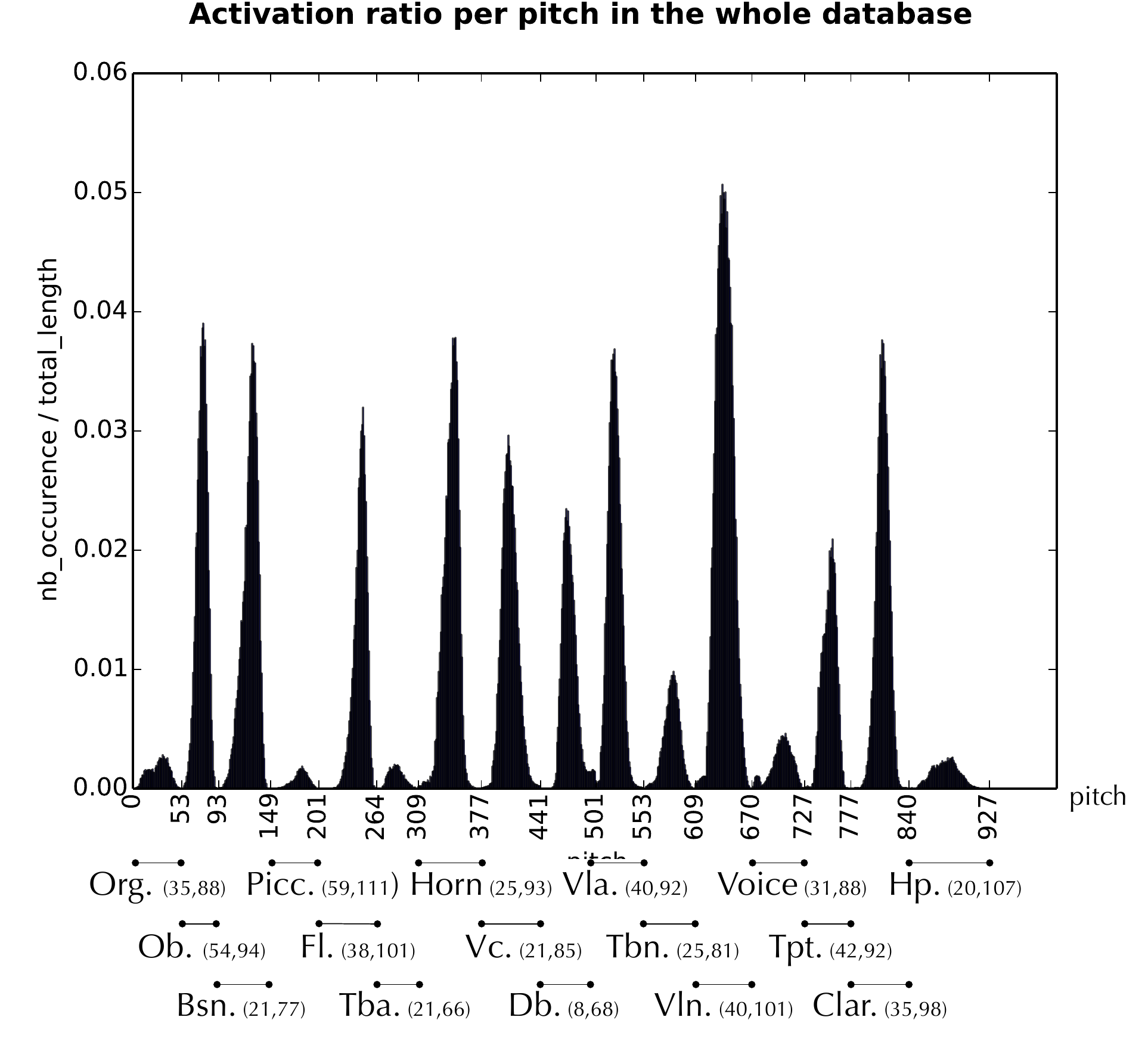}
\caption{Activation ratio per pitch in the whole orchestral score database. 
For one bin on the horizontal axis, the height of the bar represents the number of notes played by this instrument divided by the total number of frames in the database.
This value is computed for the event-level aligned representations \ref{sec:proposed_model}.
The different instruments are covered by the pitch axis, and one can observe the peaks that their medium ranges form.
The maximum value of the vertical axis (0.06), which is well below 1, indicates that each pitch is rarely played in the whole database.
}
\label{fig:pitch_histo}
\end{figure}

Figure \ref{fig:pitch_histo} highlights the activation ratio of each pitch in the orchestration scores ($\frac{\# \left\lbrace \text{pitch on} \right\rbrace}{\# \left\lbrace \text{pitch on} \right\rbrace + \# \left\lbrace \text{pitch off} \right\rbrace}$, where $\#$ is the cardinal of an ensemble) over the whole dataset. Note that this activation ratio does not take the duration of notes into consideration, but only their number of occurrences.
The pitch range of each instrument can be observed beneath the horizontal axis.

Two different kinds of imbalance can be observed in figure \ref{fig:pitch_histo}.
First, a given pitch is rarely played. Second, some pitches are played more often compared with others.
Class imbalance is known as being problematic for machine learning systems, and these two observations highlight how challenging the projective orchestration task is.
More statistics about the whole database can be found on the companion website.

\subsubsection{Integrity}
Both the metadata and instrumentation \textit{CSV} files have been automatically generated but manually checked. We followed a conservative approach by automatically rejecting any score with the slightest ambiguity between a track name and a possible instrument (for instance \textit{bass} can refer to \textit{double-bass} or \textit{voice bass}).

\subsubsection{Formats}
\begin{figure}
\centering
\includegraphics[scale=0.40]{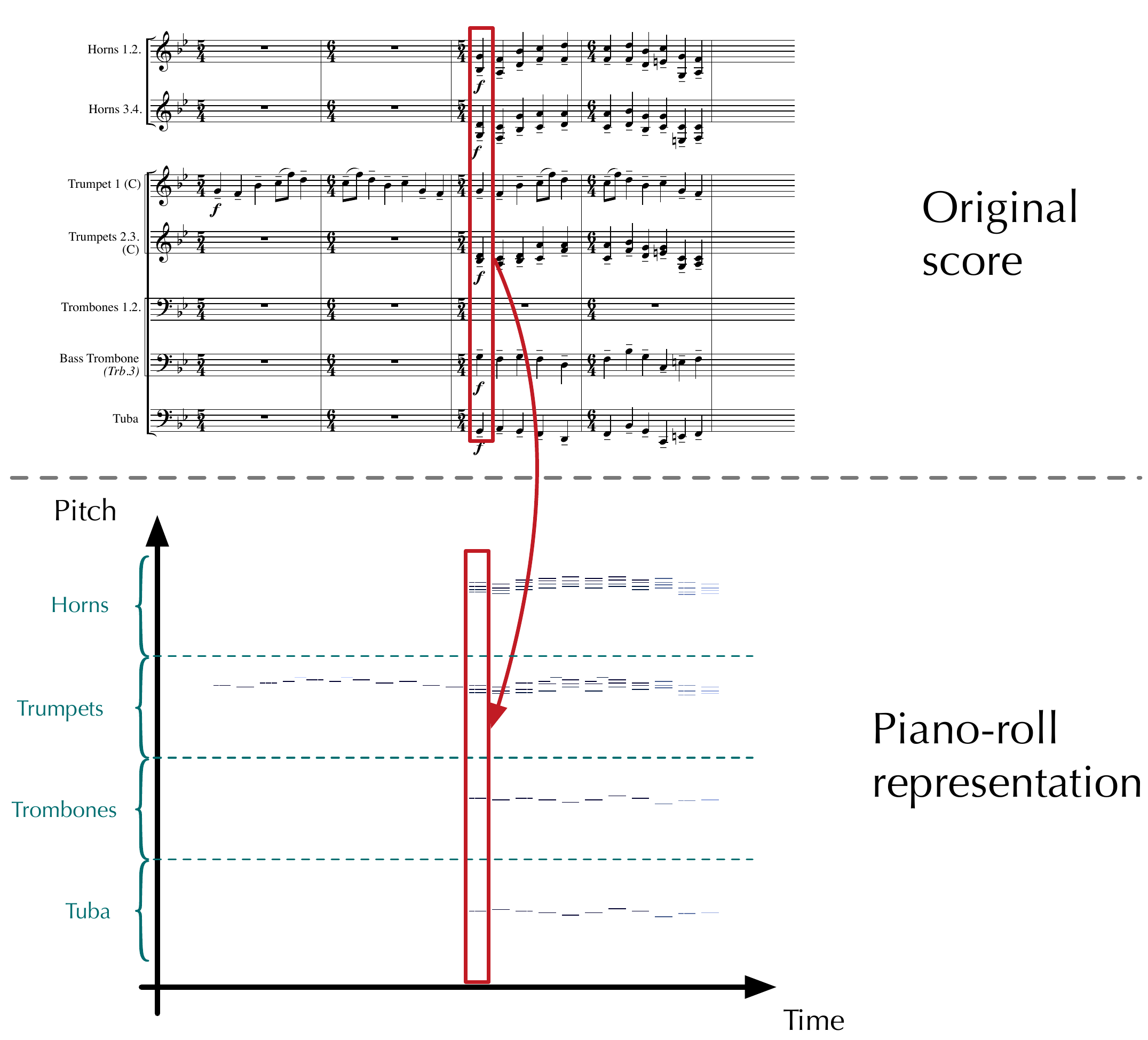}
\caption{\textit{Piano-roll} representation of orchestral scores.
The \textit{piano-roll} $pr$ is a matrix. A pitch $p$ at time $t$ played with an intensity $i$ is represented by $pr(p,t) = i$, where $0$ is a note off. This definition is extended to an orchestra by simply concatenating the \textit{piano-rolls} of every instrument along the pitch dimension.}
\label{fig:piano-roll}
\end{figure}

To facilitate the research work,
we provide pre-computed piano-roll representations such as the one displayed in \figref{fig:piano-roll}. In this case, all the \textit{MIDI} files of piano (respectively orchestra) work have been transformed and concatenated into a unique two-dimensional matrix.
The starting and ending time of each track is indicated in the \textit{metadata.pkl} file.
These matrices can be found in  Lua/Torch (.t7), Matlab (.m), Python (.npy) and raw (.csv) data formats.

\subsubsection{Score Alignment}
Two versions of the database are provided. The first version contains unmodified midi files. The second version contains \textit{MIDI} files automatically aligned using the \textit{Needleman-Wunsch} \cite{NEEDLEMAN1970443} algorithm as detailed in \secref{sec:automatic-alignment}.

\subsection{Automatic Alignment}
\label{sec:automatic-alignment}
Given the diverse origins of the \textit{MIDI} files, a piano score and its corresponding orchestration are almost never aligned temporally.
These misalignments are very problematic for learning or mining tasks, and in general for any processing which intends to take advantage of the joint information provided by the piano and orchestral scores. Hence, we propose a method to automatically align two scores, and released its Python implementation on the companion website \footnote{\url{https://qsdfo.github.io/LOP/code}}.
More precisely, we consider the piano-roll representations (\figref{fig:piano-roll}) where the scores are represented as a sequence of vectors. By defining a distance between two vectors, the problem of aligning two scores can be cast as a univariate sequence-alignment problem.

\subsubsection{Needleman-Wunsch}
The \textit{Needleman-Wunsch} (\textit{NW}) algorithm \cite{NEEDLEMAN1970443} is a dynamic programming technique, which finds the optimal alignment between two symbolic sequences by allowing the introduction of gaps (empty spaces) in the sequences.
An application of the \textit{NW} algorithm to the automatic alignment of musical performances is introduced in \cite{grachten2013automatic}.
As pointed out in that article, \textit{NW} is the most adapted technique for aligning two sequences with important structural differences like skipped parts, for instance.

The application of the \textit{NW} algorithm relies solely on the definition of a cost function, which allows the pairwise comparison of elements from the two sequences, and the cost of opening or extending a gap in one of the two sequences.

\subsubsection{Similarity Function}
To measure the similarity between two chords, we propose the following process:
\begin{itemize}
\item discard intensities by representing notes being played as one and zero otherwise.
\item compute the pitch-class representation of the two vectors, which flattens all notes to a single octave vector (12 notes).
In our case, we set the pitch-class to one if at least one note of the class is played.
For instance, we set the pitch-class of C to one if there is any note with pitch C played in the piano-roll vector.
This provides an extremely rough approximation of the harmony, which proved to be sufficient for aligning two scores.
After this step, the dimensions of each vector is 12.
\item if one of the vectors is only filled with zeros, it represents a silence, and the similarity is automatically set to zero (note that the score function can take negative values).	
\item for two pitch-class vectors $\bm{A}$ and $\bm{B}$, we define the score as 
\begin{equation}
S(\bm{A}, \bm{B}) \ = C \ \times \ \frac{\sum_{i=1}^{12} \delta(A_i + B_i)}{\ max(||\bm{A}+\bm{B}||_1 , 1)}
\label{eq:score_function}
\end{equation}
where $\delta$ is defined as:
\[
    \delta(x) = \left\{
     	\begin{array}{rl}
        0 & \text{if x = 0}\\
        -1 & \text{if x = 1}\\
     	1 & \text{if x = 2}
        \end{array} \right.
\]
$C$ is a tunable parameter
and $||x||_{1} = \sum_{i} |x_i|$ is the $\mathcal{L}1$ norm.
\end{itemize}
Based on the values recommended in \cite{NEEDLEMAN1970443} and our own experimentations, we set C to 10.
The gap-open parameter, which defines the cost of introducing a gap in one of the two sequences, is set to 3 and the gap-extend parameter, which defines the cost of extending a gap in one of the two sequences, is set to 1.

\section{An application : projective automatic orchestration}
In this section, we introduce and formalize the automatic projective orchestration task (\figref{fig:orch}). In particular, we propose a system based on statistical learning and define an evaluation framework for using the \textit{POD} database.
 
\subsection{Task Definition}
\label{sec:task_definition}
\subsubsection{Orchestral Inference}
For each orchestral piece, we define as $\textbf{O}$ and $\textbf{P}$ the aligned sequences of column vectors from the \textit{piano-roll} of the orchestra and piano parts.
We denote as $T$ the length of the aligned sequences $\textbf{O}$ and $\textbf{P}$.

The objective of this task is to infer the present orchestral frame knowing both the recent past of the orchestra sequence and the present piano frame. Mathematically, it consists in designing a function $f$ where
\begin{equation}
\hat{O}(t) = \bm{f}\lbrack P(t), O(t-1), ..., O(t-N) \rbrack \quad \forall t \in \left\lbrack N, ... T \right\rbrack
\label{eq:inference_function}
\end{equation}
and $N$ defines the order of the model.

\subsubsection{Evaluation Framework}
We propose a quantitative evaluation framework based on a one-step predictive task.
As discussed in \cite{conklin1995multiple}, we make the assumption that an accurate predictive model will be able to generate original acceptable works.
Whereas evaluating the generation of a complete musical score is subjective and difficult to quantify, a predictive framework provides us with a quantitative evaluation of the performance of a model.
Indeed, many satisfying orchestrations can be created from the same piano score.
However, the number of reasonable inferences of an orchestral frame given its context (as described in equation \ref{eq:inference_function}) is much more limited.

As suggested in \cite{boulanger2012modeling,lavrenko2003polyphonic}, the accuracy measure \cite{bay2009evaluation} can be used to compare an inferred frame $\hat{O}(t)$ drawn from \eqref{eq:inference_function} to the ground-truth $O(t)$ from the original file.
\begin{equation}
\text{Accuracy}(t)  = 100 \, . \, \frac{TP(t)}{TP(t) + FP(t) + FN(t)}
\label{eq:accuracy}
\end{equation}
where $TP(t)$ (true positives) is the number of notes correctly predicted (note played in both $\hat{O}(t)$ and $O(t)$). $FP(t)$ (false positive) is the number of notes predicted that are not in the original sequence (note played in $\hat{O}(t)$ but not in $O(t)$). $FN(t)$ (false negative) is the number of unreported notes (note absent in $\hat{O}(t)$, but played in $O(t)$).


When the quantization gets finer, we observed that a model which simply repeats the previous frame gradually obtains the best accuracy as displayed in \tabref{tab:results}.
To correct this bias, we recommend using an event-level evaluation framework where the comparisons between the ground truth and the model's output is only performed for time indices in $T_e$ defined as the set of indexes $t_{e}$ such that
\[\text{O}(t_{e}) \neq \text{O}(t_{e} - 1)\]
The definition of event-level indices can be observed in \figref{fig:event_level_generation}.

In the context of learning algorithms, splitting the database between disjoint \textit{train} and \textit{test} subsets is highly recommended \cite[pg.32-33]{bishop2006pattern}, and the performance of a given model is only assessed on the test subset.
Finally, the mean accuracy measure over the dataset is given by
\begin{equation}
\frac{1}{K} \sum_{s \in \mathcal{D}_{test}} \sum_{t_e \in T_e(s)} Accuracy(t_e)
\end{equation}
where $\mathcal{D}_{test}$ defines the test subset, $T_{e}(s)$ the set of event-time indexes for a given score s, and $K = \sum_{s \in \mathcal{D}_{test}} |T_e(s)|$.

\subsection{Proposed Model}
\label{sec:proposed_model}
In this section, we propose a learning-based approach to tackle the automatic orchestral inference task.

\subsubsection{Models}
We present the results for two models called \textit{conditional Restricted Boltzmann Machine} (\textit{cRBM}) and \textit{Factored Gated cRBM} (\textit{FGcRBM}).
The models we explored are defined in a probabilistic framework, where the vectors $O(t)$ and $P(t)$ are represented as binary random variables. The orchestral inference function is a neural network that expresses the conditional dependencies between the different variables: the present orchestral frame $O(t)$, the present piano frame $P(t)$ and the past orchestral frames $O(t-1,...,t-N)$. \textit{Hidden units} are introduced to model the co-activation of these variables. Their number is a hyper-parameter with an order of magnitude of 1000.
A theoretical introduction to these models can be found in \cite{taylor2009factored}, whereas their application to projective orchestration is detailed in \cite{lop_smc}.

\subsubsection{Data Representation}
In order to process the scores, we import them as \textit{piano-roll} matrices (see \figref{fig:piano-roll}). 
Their extension to orchestral scores is obtained by concatenating the \textit{piano-rolls} of each instrument along the pitch dimension.

\begin{figure}
\centering
\includegraphics[scale=0.15]{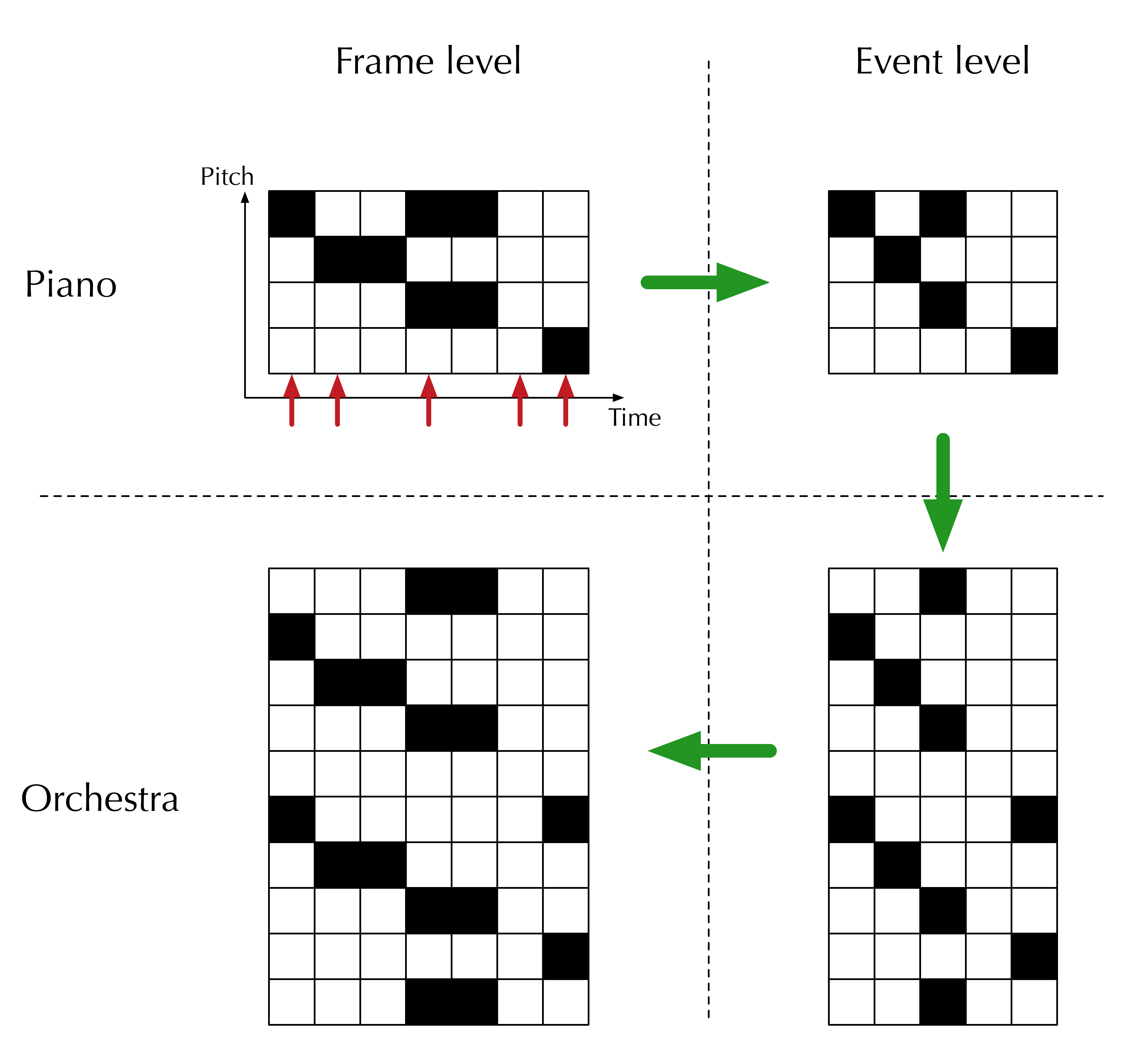}
\caption{From a piano score, the generation of an orchestral score consists in extracting the event-level representation of the piano score, generating the sequence of orchestral events, and then injecting them at the position of the event from the piano score. Note that the silence in the fourth event of the piano score is not orchestrated by the probabilistic model, but is automatically mapped to a silence in the orchestral version.}
\label{fig:event_level_generation}
\end{figure}

Then, new events $t_e \in T_e$ are extracted from both piano-rolls as described in \secref{sec:task_definition}.
A consequence is that the trained model apprehends the scores as a succession of events with no rhythmic structure.
This is a simplification that considers  the rhythmic structure of the projected orchestral score to be exactly the same as the one of the original piano score. This is false in the general case, since a composer can decide to add nonexistent events in an orchestration.
However, this provides a reasonable approximation that is verified in a vast majority of cases.
During the generation of an orchestral score given a piano score, the next orchestral frame is predicted in the event-level framework, but inserted at the temporal location of the corresponding piano frame as depicted in \figref{fig:event_level_generation}.

Automatic alignment of the two \textit{piano-rolls} is performed on the event-level representations, as described in \secref{sec:automatic-alignment}.

In order to reduce the input dimensionality, we systematically remove any pitch which is never played in the training database for each instrument. With that simplification the dimension of the orchestral vector typically decreases from 3584 to 795 and the piano vector dimension from 128 to 89.
Also, we follow the usual orchestral simplifications used when writing orchestral scores by grouping together all the instruments of a same section. For instance, the \textit{violin} section, which might be composed by several instrumentalists, is written as a single part.
Finally, the velocity information is discarded, since we use binary units that solely indicate if a note is on or off.

Eventually, we observed that an important proportion of the frames are silences, which mathematically corresponds to a column vector filled with zeros in the piano-roll representation.
A consequence of the over-representation of silences is that a model trained on this database will lean towards orchestrating with a silence any piano input, which is statistically the most relevant choice.
Therefore, orchestration of silences in the piano score ($P(t) = 0$) are not used as training points.
However, it is important to note that they are not removed from the piano-rolls.
Hence, silences could still appear in the past sequence of a training point, since it is a valuable information regarding the structure of the piece.
During generation time, the silences in the piano score are automatically orchestrated with a silence in the orchestra score. Besides, silences are taken into consideration when computing the accuracy.

\subsubsection{Results}
The results of the \textit{cRBM} and \textit{FGcRBM} on the orchestral inference task are compared to two na\"ive models. The first model is a random generation of the orchestral frames obtained by sampling a Bernoulli distribution of parameter $0.5$. The second model predicts an orchestral frame at time $t$ by simply repeating the frame at time $t-1$.
The results are summed up in \tabref{tab:results}.

\subsection{Discussion}
As expected, the random model obtains very poor results.
The repeat model outperform all three other models, surprisingly even in the event-level framework.
Indeed, we observed that repeated notes  still occur frequently in the event-level framework.
For instance, if between two successive events only one note out of five is modified, the accuracy of the repeat model on this frame will be equal to $66\%$.

If the \textit{FGcRBM} model outperforms the \textit{cRBM} model in the frame-level framework, the \textit{cRBM} is slightly better than the \textit{FGcRBM} model in the event-level framework.

Generations from both models can be listened to on the companion website \footnote{\url{https://qsdfo.github.io/LOP/results}}.
Even though some fragments are coherent regarding the piano score and the recent past orchestration, the results are mostly unsatisfying.
Indeed, we observed that the models learn an extremely high probability for every note to be off.
Using regularization methods such as weight decay has not proven efficient.
We believe that this is due to the sparsity of the vectors $O(t)$ we try to generate, and finding a more adapted data representation of the input will be a crucial step.

\section{Conclusion and future work}
We introduced the Projective Orchestral Database (\textit{POD}), a collection of \textit{MIDI} files dedicated to the study of the relations between piano scores and corresponding orchestrations. 
We believe that the recent advent in machine learning and data mining has provided the proper tools to take advantage of this important mass of information and investigate the correlations between a piano score and its orchestrations.
We provide all \textit{MIDI} files freely, along with aligned and non-aligned pre-processed piano-roll representations on the website \url{https://qsdfo.github.io/LOP/index.html}.

We proposed a task called automatic orchestral inference. Given a piano score and a corresponding orchestration, it consists in trying to predict orchestral time frames, knowing the corresponding piano frame and the recent past of the orchestra. 
Then, we introduced an evaluation framework for this task based on a train and test split of the database, and the definition of an accuracy measure.
We finally present the results of two models (the \textit{cRBM} and \textit{FGcRBM}) in this framework.

We hope that the \textit{POD} will be useful for many researchers.
Besides the projective orchestration task we defined in this article, the database can be used in several other applications, such as generating data for a source-separation model \cite{3765}.
Even if  small errors still persist, we thoroughly checked manually the database and guarantee its good quality.
However, the number of files collected is still small with the aim of leading statistical investigations.
Hence, we also hope that people will contribute to enlarge this database by sharing files and helping us gather the missing information.

\bibliography{biblio}

%
%
%
%

\end{document}